\documentclass{aa}
\usepackage{psfig}
\def\J{\scriptscriptstyle{J}}
\def\IC{\scriptscriptstyle{IC}}
\def\smalltimes{\;{\scriptstyle{\times}}}
\newcommand{\siml} {\lower.5ex\hbox{$\; \buildrel < \over \sim \;$}}
\newcommand{\simg} {\lower.5ex\hbox{$\; \buildrel > \over \sim \;$}}

\begin{document}
	
	\thesaurus{02     
              (02.18.5;  
               13.07.1;  
	       13.07.3;  
               02.18.8)} 
   \title{Spectral Evolution of GRBs and the ``Death Line'' of the Synchrotron Shock Model}

   \author{H. Papathanassiou}

   \offprints{H. Papathanassiou}

   \institute{International School for Advanced Studies (SISSA),
		Via Beirut 2-4, 34013 Trieste, ITALY\\
              email:  hara@sissa.it
             }

   \date{Received January 22, 1999; accepted March 10, 1999}

   \maketitle

   \begin{abstract}
I calculate spectral evolution series for pulses of GRBs,
in  the BATSE spectral range, for continuous particle injection and cooling by
synchrotron, inverse Compton, and adiabatic expansion. 
The hydrodynamic properties of the relativistic outflow are 
homogeneous across the emitting region which is a conical jet. The flow is viewed at an angle 
off its symmetry axis; time delays are taken into account. 
I discuss the low energy slope part of the spectrum, in view of the 
recent claim of photon slopes in bright BATSE bursts that are inconsistent with the optically thin synchrotron shock model (SSM).
   
   \end{abstract}

%

\section{Introduction}
The nice agreement between the spectral evolution of afterglows and an
adiabatically and relativistically expanding spherical shell that emits through
synchrotron ({\it sy}) has established the SSM for the description of the late stages of the evolution of the fireball that is believed to
give rise to the main GRB event and its afterglow (e.g., 
\cite{Wijers97}, \cite{Galama}).
These considerations imply that {\it sy} is the dominant cooling mechanism
for the electrons of the flow during at least the afterglow phase and therefore
the magnetic field $B$ is close to its equipartition value (see also \cite{WijersGalama}).
On the other hand, spectral fitting of time resolved BATSE spectra of bright GRBs
 (\cite{Preece98}) has yielded a significant number of cases with low energy photon  indices $\alpha$ ($n_{\nu} \sim \nu^{-\alpha}$) exceeding -2/3; this is inconsistent with
the SSM and $\alpha =-2/3$ has been referred to as its ``death line''.
One way for the SSM model to overcome this difficulty
is if radiation  became self absorbed in the BATSE window for a portion of the 
burst.

Under the reasonable extrapolation that the MeV range spectrum of the GRB proper is also due to the {\it sy} component, I calculate observed 
spectral evolution series for GRB pulses. 
The relativistic motion of the region may cause the observed spectra to have
a different {\it shape} from the intrinsic (``co-moving'').

I use values that are appropriate for the description of internal shocks
(\cite{MR_internal}) and 
where the $e^-$ come from the ionization of the baryonic material.
Equipartition values of the physical parameters for a flow of total luminosity per unit solid angle
$L= 10^{52} L_{52} $ erg/s sterad, expanding at constant Lorenz factor $\Gamma =300$
and with intrinsic variability timescale $t_{var}$ are $B\simeq 2\smalltimes 10^3 \sqrt{L_{52}}/t_{var}$~G and $n_e \simeq n_p \simeq 10^8 L_{52}/t_{var}^2 \,\rm{cm}^{-3}$. 
I present broad band spectra and discuss the effect that pairs may have on the
BATSE component.

\section{Set up and Calculation}

The emitting region consists of a conical shell of opening angle $2\Theta_{\J}$,
the base of which is at $r_b = r_o +\beta t$ (where $\beta=\sqrt{1 -1/\Gamma^2}, \Gamma\gg 1$, and  $r_o =\Gamma^2 c t_{var}$ is the dissipation radius) and
its front is moving at 
$\beta_{sh,f}$($\sim {\cal{O}}(10^{-1}))$ in the flow's frame.
 The emitting region is assumed homogeneous at any given moment. This is valid 
as long as the region is not expanding faster than the injected electrons can fill
it up (i.e., $\beta_{sh,f} \ll 1$).
$e^-$s are injected at a constant rate per unit volume with the same spectral shape, i.e., a relativistic Maxwellian peaking at
$\gamma_{m,o}$ and a power law tail of slope $-3$ extending up to $\gamma_{M,o}$ (determined by the size of the emitting region).
I solve the continuity equation along its characteristics. I take into account 
continuous particle injection (over $t_{inj}$),
cooling through {\it sy} ($|d\gamma/dt|_{sy} = 1.29 \smalltimes 10^{-9} B^2 (\gamma^2$ -1)), IC ($|d\gamma/dt|_{\IC}=  1.36\smalltimes 10^{-17}(\gamma^2-1) \int_0^{m_e c^2/h \gamma}{\cal{I}}^{'}_{sy,\nu'} d\nu'$ and the {\it sy} intensity integral is evaluated iteratively so that IC cooling is taken into account self-consistently) and   adiabatic expansion of the shell $|d\gamma/dt|_{ad}= \gamma(dn_p/n_p +\beta/(r_o +\beta t))$.
The observer lies at an angle $\Theta_o$ from the symmetry axis of the jet. Thus,
the frequency of a photon emitted at a angle $\theta$ with respect to the jet
axis  is 
boosted by a Doppler factor ${\cal{D}}= \left\{\Gamma \left[1- \beta cos(\theta+\Theta_o)\right]\right\}^{-1}$. The observed spectrum is calculated by integrating the contributions from each volume element over the visible area and the duration of the emission.

For typical values of the physical parameters, the BATSE range spectrum is entirely optically thin.
The different scenarios examined include the following:
Continuous injection in a thin shell (curvature determines the lightcurve); Continuous injection in a thick shell (thickness determines lightcurve);
Magnetic field decay with the regions expansion (conserving either flux or total energy); Smooth time dependence of injected $e^-$ number (conserving either total number or number density); 
The jet is viewed off the axis of symmetry. 

\section{Results and Discussion}

As the {\it instantaneous} observed spectrum results from sampling of the contributions
from  different parts of the flow emitted at different times in the flow's 
evolution its {shape} may be different from that of the intrinsic one.
For example, the
expansion of the emitting region (or, alternatively, the increase of the $e^-$ number density) makes  the comoving self-absorption 
frequency increase with time which, in turn, results in spectral slopes of the observed spectra that are flatter by 0.5 (i.e., the portion of slope 2 (2.5) in 
spectral flux of the co-moving spectrum is mapped to a slope 1.5 (2) part in the observed spectrum).
None of the scenarios examined here can cause
the low energy photon number slope to exceed the ``death line'' limit.

Consider first the simplest approach where the $e^-$s are injected instantaneously and fill up the whole volume. For the self-absorption frequency to fall in the BATSE range and the peak $e^-$s' radiation be self-absorbed, the
physical parameters have to take values (in {\it cgs} units) in the following range: $10^2 \siml B \siml 10^{3.5}$ and $ 10^{15} \siml 
n_e \siml 10^{18.5}$ for $\gamma_{m,o} \simeq 10^3$, while $n_e$ can be low ($ 10^{10} \siml n_e \siml 10^{15}$) provided
$10^9 \simg B \simg 10^{7}$ for $\gamma_{m,o} \simeq 2$. In all cases, at least
one of the parameters has to take values that are substantially higher than the
equipartition ones, in the framework of dissipative flows.

\begin{figure}
	\vspace{1cm}
	\hspace{0.5cm}\psfig{figure=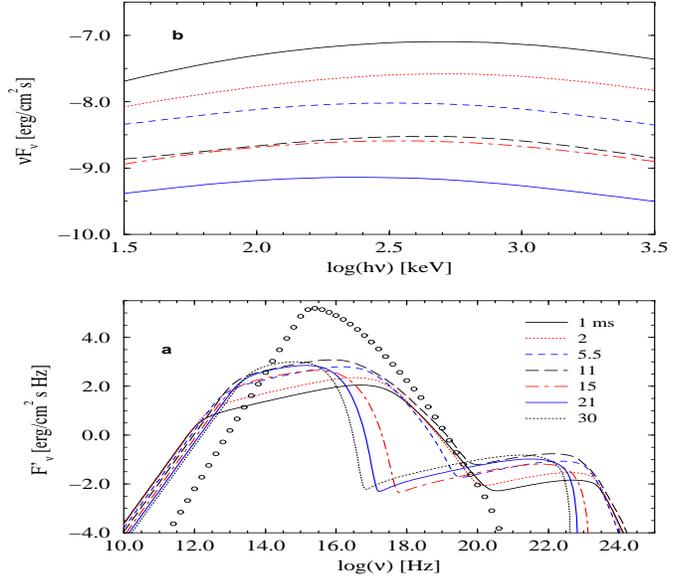,width=8.8cm,height=6.5cm}
	\vspace{0cm}
      \caption[]{Spectral evolution series. {\bf{a}} Broad band comoving spectra. {\bf{b}} Time resolved (sampled over $\Delta T_{det}=10$ ms every 10 ms) spectra in the BATSE window.
              }
         \label{fig:equip}
   \end{figure}

In Fig.~\ref{fig:equip}, I present a sequence of time resolved spectra for
a flow of $L_{52}=1$, $t_{var}= 0.1$~s, $\Theta_o =0^o$, $n_e$ and $B$ equal to the equipartition values, $\gamma_{m,o} =3000$, and $t_{inj}=t_{var}/10$. This set of
parameters will favor high values of the self-absorption frequency for a relativistic flow that develops internal shocks.
The lower panel shows the instantaneous spectra in the fluid frame in $F_{\nu} $ vs $\nu$. During injection, there is brightening and progression of the optically thick part to harder frequencies while, after that, rapid softening takes place.
The upper panel shows the observed BATSE spectrum. The pulse is detectable for 60~ms during which time it is dimming always retaining the typical {\it sy} 
slopes.

One way to circumvent the problem of the high values of the slopes is to increase the $e^-$ content of the flow.
This might happen if the flow has a high compactness and produces a
large density of pairs that live long enough to contribute to the {\it sy}
emission and turn it optically thick in the BATSE range.
Pilla \& Loeb (1997) have stressed the importance of the pairs in internal shocks,
although they do not calculate any effect these might have on the optical depth.
To assess the importance of pairs in the flow, I include, in the lower panel of Fig.~\ref{fig:equip}, the spectrum of the first snapshot as this is modified by the pairs that result from the absorption of the IC photons. 
At this time, all the hard photons above 1 MeV are absorbed (which is consistent with the limits on the hard GRB counterparts) 
and some $10^{13} 
\rm{cm}^{-3}$ pairs with $\gamma \simg 1$ and a power law distribution fill up
the region resulting into 
 brightening and steepening of the soft part of the {\it sy} component. 
Their annihilation timescale is of the order of s and they will cool mainly through IC.
I stress that this is preliminary only, and one has to include the pairs in 
the emitting population in a self consistent way.
Therefore, while  {\it sy}  is responsible for the BATSE component, in flows with high $n_e$ and thus bright IC component,  pairs might be able to provide the required opacity for the {\it sy} component to become self-absorbed in a transient fashion. For high pair production rates, the self-absorption peak can fall in
the BATSE window. 
%

\begin{acknowledgements}
I thank Ph. Papadopoulos, A. Celotti and P. M\'esz\'aros for useful comments.
      This work was supported by the Italian \emph{MURST}.
\end{acknowledgements}

\end{document}